\documentclass[aps,prl,twocolumn,groupedaddress,showpacs,preprintnumbers,amsmath,amssymb]{revtex4}
\usepackage{graphics,graphicx} % EPS

\newcommand{\meanval}[1]{\left\langle #1 \right\rangle} % <A>
\newcommand{\InsertFig}[1]{\includegraphics{#1.eps}} % EPS (pst2eps script)
\newcommand{\setZ}{{\mathord{\mathbb Z}}}

\begin{document}

\preprint{LPTM/2004-REY02}

\title{Simulation of Spin Models in Multicanonical Ensemble with Collective Updates}

\author{S. Reynal}
\email{reynal@ensea.fr}
\homepage{http://www-reynal.ensea.fr}
\altaffiliation{Permanent address: ENSEA, 6 Av. du Ponceau, 95014 Cergy Cedex, France.}
\author{H. T. Diep}
\affiliation{Laboratoire de Physique Th\'eorique et Mod\'elisation, CNRS-Universit\'e de Cergy-Pontoise, 5 mail Gay-Lussac, Neuville sur Oise, 95031 Cergy-Pontoise Cedex, France}

\date{\today}

\begin{abstract}
We propose a Monte Carlo method which performs a random walk in energy space using cluster-like collective updates.
By imposing that bond probabilities depend continuously on the microcanonical temperature, we obtain dynamic exponents close to their ideal random walk values. The method  proves remarkably powerful when applied to models governed by long-range interactions, where it straightforwardly combines with the efficient Luijten-Bl\"ote cluster algorithm to yield a dramatic reduction in the computation load. 
\end{abstract}

\maketitle

Of the many methods dedicated to the study of spin models, Monte Carlo (MC) methods have now gained a prominent role.
As regards models exhibiting either first-order transitions or disorder-induced rough free energy landscapes, {\em  canonical} MC simulations are known to suffer, however, from supercritical slowing down \cite{Berg1992}, an exponential growth with the lattice size of the tunneling time between free energy minima that leads to unreliable statistics.
An efficient approach aimed at beating this limitation is the simulation in generalized ensembles \cite{Iba2001,Berg2002}, in particular its multicanonical flavor initially proposed by Berg \cite{BergNeuhaus1992,Berg1992} and independently by Lee \cite{Lee1993}, reconsidered in the framework of transition matrix dynamics \cite{Wang1999}, and recently revisited by Wang and Landau \cite{WangLandau2001b}. The key-idea here is to artificially enhance
 rare events corresponding to local maxima in the free energy, by feeding the Markovian chain with an auxiliary distribution $W(E)$ best approximating the inverse of the density of states. 
Indeed, this was shown to reduce tunneling times from an exponential to a power law $\tau \sim L^z$ of the lattice size \cite{BergNeuhaus1992}.
Still, simulations in the multicanonical ensemble based on local updates yield dynamic exponents $z$ which are substantially higher than their ideal random walk value $z \sim D$. In the case of long-ranged (LR) models, there is an additional hurdle owing to the very presence of long-range interactions which makes the computation of the energy --- an essential ingredient of multicanonical algorithms --- a very time consuming operation, namely, of $O(L^2)$ complexity \cite{ReynalDiep2004a}.

In this Letter, we propose a Monte Carlo method which successfully tackles these two issues by performing simulations in the multicanonical ensemble using collective updates. 
As opposed to previous approaches aimed at combining cluster updates with a multicanonical algorithm
 \cite{Rummukainen1993,JankeKappler1995b,Yamaguchi2003},
our method relies on a straightforward cluster-building mechanism which hinges on the microcanonical temperature of the current configuration in order to determine appropriate bond probabilities. 
We test the efficiency of the method on the two-dimensional $q$-state Potts model with nearest-neighbor (NN) interactions ($q=7,10$) and on its one-dimensional LR counterpart with $1/r^{1+\sigma}$ interactions ($q=3,6,12$), with parameters chosen so that a first-order regime is exhibited. In both test cases, analyses of tunneling rates show a very substantial reduction in the dynamic exponents, from e.g., $z=1.35(3)$ to $z=1.05(1)$ for the LR model with $q=6$ and $\sigma=0.7$, and from $z=2.60(4)$ to $z=1.82(2)$ for the two-dimensional NN model with $q=7$.
We further demonstrate that our formulation makes it exceptionally straightforward to incorporate two acceleration schemes dedicated to LR models \cite{Luijten1995,KrechLuijten2000}, which cut down the algorithm complexity from $O(L^2)$ to $O(L\ln(L))$. Chains containing up to $2^{16}$ spins were simulated in a few days, whereas challenging such huge sizes with local updates would have demanded several months of intensive computation. 

\paragraph{Algorithm}

In the multicanonical method, one wishes to sample a flat histogram of the energy over a given energy range. The weight $w(E)$ of a state of energy $E$ is thus set to the inverse of the density of states, or more specifically, to an estimate of it  obtained, in our case, using the Wang-Landau method \cite{WangLandau2001b}. Denoting the microcanonical entropy as $S(E)$, one may write $w(E)=e^{-S(E)}$. A local-update algorithm consists in updating a single spin and accepting the attempted move with a probability given by $\min \left[1, e^{S(E_a)-S(E_b)}\right]$, where $E_a$ and $E_b$ stand for the energy of the initial and final states, respectively. In order to allow this algorithm to embody a collective-update scheme, we first rewrite the multicanonical weight $w(E)$ as $e^{-\beta(E) E} \phi(E)$ where $\beta(E) = dS(E)/dE$ is the inverse microcanonical temperature. The very presence of a term having the same form as the canonical Boltzman weight provides the means to reexpress the multicanonical weight as a trace over the bonds of a Fortuin-Kasteleyn random cluster \cite{FortuinKasteleyn1969}, thus paving the way for a collective-update scheme. 
Although our algorithm may be equally well applied to other spin models, e.g., models incorporating disorder 
or exhibiting a continuous symmetry,
we now consider, for the sake of clarity, a generalized ferromagnetic spin model with a $\setZ_q$ symmetry, whose energy reads $E = - \sum_{i < j} J(|i-j|) \delta_{\sigma_i,\sigma_j}$. Here $J(|i-j|)>0$ and the $\sigma_i$ variables can take on integer values between $1$ and $q$. 
Invoking the Fortuin-Kasteleyn representation of this model, we can write the multicanonical weight $w(E)$ as 
$$
	w(E) = \phi(E) \sum_{[b]} \prod_{i<j} p_{|i-j|}(E) \delta_{\sigma_i,\sigma_j} \delta_{b_{ij},1} + \delta_{b_{ij},0}.
$$
where the sum runs over all lattice bonds, a bond is active (inactive) whenever $b_{ij}=1$ ($0$),
and $p_{|i-j|}(E)=e^{\beta(E) J(|i-j|)} - 1$.
A collective-update step consists of two stages, namely first building a set of clusters from the current configuration, and then updating all clusters at once with an acceptance probability which ensures that detailed balance is preserved.
The cluster construction can be carried out in exactly the same way as in the original Swendsen-Wang algorithm \cite{SwendsenWang1987}. Starting from a configuration at energy $E_a$ and an empty bond set, we consider each pair of identical spins $\{\sigma_i,\sigma_j\}$ in turn, and place a bond with probability $1-e^{-\beta(E_a) J(|i-j|)}$. We then identify clusters of connected spins, and draw a new spin value at random for each cluster.
Observing that a given bond configuration at energy $E$ has a weight $\phi(E) \prod_{l>0} p_l(E)^{B(l)}$, where $B(l)$ is the number of bonds of length $l=|i-j|$, we therefore accept the attempted cluster flips with the following acceptance rate:
\begin{equation}
	W_{flip}(a\rightarrow b) = \min \left\{1, \frac{\phi(E_b)}{\phi(E_a)} \prod_{l>0} \left[ \frac{p_l(E_b)}{p_l(E_a)}\right]^{B(l)} \right\}
\label{eq:cluster_flip_acceptance_rate}
\end{equation}
Here, $E_b$ denotes the energy of the new spin configuration. Since this acceptance rate is nothing but that of a Metropolis algorithm, detailed balance is trivially satisfied. In particular, if we consider a canonical simulation at inverse temperature $\beta_0$, we have $\beta(E)=\beta_0$ and $\phi(E)=1$; hence the acceptance rate is equal to $1$, and we are back to the original Swendsen-Wang algorithm.
It is crucial to underline that it is the microcanonical temperature, or equivalently the lattice energy, which entirely governs the cluster construction; indeed, for a given lattice configuration at energy $E$, bonds are placed as if the model were simulated at its micro-canonical temperature using a Swendsen-Wang algorithm. As a result, cluster bond probabilities change continuously as the lattice configuration walks along the available energy range of the random walk, so that, e.g., small clusters are built in the upper energy range and conversely large clusters in the lower energy range. Obviously, this mechanism entails determining $\beta(E)$ to sufficient accuracy (any departure from the ideal line resulting in poorer performances). We chose to compute $\beta(E)$ from the estimated density of states using spline interpolations, yet other means are available, e.g., one may rely on transition matrices \cite{Swendsen2004}; this approach, though being slower, gave smoother estimates already at the very beginning of the simulation.

We now examine two optimization schemes suited for long-ranged models.
According to Eq.~\ref{eq:cluster_flip_acceptance_rate}, determining the acceptance rate of a cluster flip demands that we compute the energy of the new (attempted) lattice configuration beforehand. For long-ranged models, this represents an $O(L^2)$ operation, yet updating the lattice configuration in a \textit{collective} way allows us to cut this complexity down to an $O(L \ln L)$ one by relying on an FFT implementation of the convolution theorem \cite{KrechLuijten2000,ReynalDiep2004d}. Still, it is crucial to note that this reduction is absolutely intractable with single-spin update implementations owing to the very reason that a single spin is updated at a time.
For long-ranged spin models, the cluster-building process represents another exceedingly time-consuming operation, since  at each MC step approximately $L^2$ pairs of spin are considered in turn for bond activation.
When interactions decay with distance, a significant amount of time during the cluster construction is further wasted because an overwhelming number of bonds considered for activation have only a negligible probability to be activated. Our formulation of the multicanonical weight $w(E)$ in terms of a Fortuin-Kasteleyn mapping makes it straightforward, however, to build clusters using the efficient Luijten-Bl\"ote method based on cumulative bond probabilities \cite{Luijten1995} whereby, instead of considering each spin in turn for addition to a given cluster, it is the index of the next spin to be added which is drawn at random. The efficiency of this algorithm does not depend on the number of interactions per spin, and leads to a CPU demand which scales roughly as $L$. 
Our implementation differs with that of \cite{Luijten1995} essentially in that the cumulative bond probabilities now depend on the lattice energy through the microcanonical temperature, which is obviously constant over the whole cluster-construction process \cite{ReynalDiep2004d}. 

\begin{figure}
	\centering
	\InsertFig{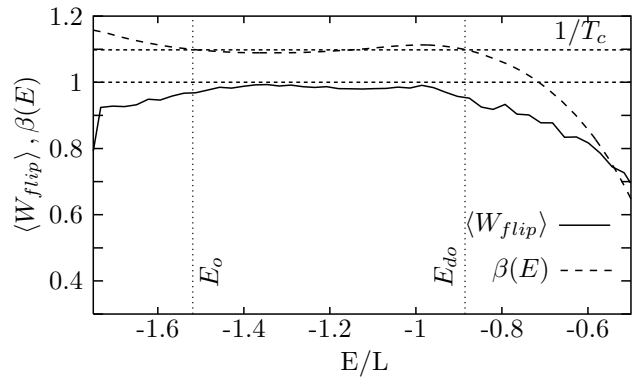}
	\caption{Mean acceptance rate $\meanval{W_{flip}}$ as a function of the energy per spin and inverse microcanonical temperature $\beta(E)$ for the long-ranged six-state Potts chain model with $\sigma=0.7$, and $L=1024$ spins. $E_o$ and $E_{do}$ denote the energy of the histogram peaks corresponding to the ordered and disordered phase, respectively. The finite size transition temperature $T_c$ and the $100\%$ line are shown for convenience.}\label{fig:mean_acceptance_rate}
\end{figure}
\paragraph{Numerical results}
We first discuss performance issues related to mean acceptance rates and tunneling times, for both the NN and the LR models.
Indeed, as opposed to the Swendsen-Wang cluster algorithm, the acceptance rate of our algorithm (Eq.~\ref{eq:cluster_flip_acceptance_rate}) is not trivially equal to unity, yet is tightly related to the efficiency with which the Markovian chain generates roughly independent samples. An approximate analytical expression of the acceptance rate when the initial and the final energies $E_a$ and $E_b$ differ only by a small amount $\epsilon$ is given to first order in $\epsilon$ by $W_{flip} = \min \left(1, 1+ \Delta(E_a) \epsilon\right)$, with 
$$
	\Delta(E_a) = \beta'(E_a) \left[ \sum_{l>0} B(l) J(l) \frac{1+p_l(E_a)}{p_l(E_a)} + E_a \right].
$$
The average energy is related to the average number of bonds of length $l$ by $\meanval{E} = -\sum_{l>0} J(l) \frac{1+p_l(E)}{p_l(E)} \meanval{B(l)}$, which shows that $\meanval{\Delta(E)}=0$. Assuming a gaussian distribution for $\Delta(E)$, it is easy to show that the variance $\meanval{\Delta(E)^2}$ is proportional to $\beta'(E)^2$ and a term varying smoothly with $E$, whence $1-\meanval{W_{flip}} \propto |\beta'(E)| \epsilon$.
As shown in Fig.~\ref{fig:mean_acceptance_rate}, our numerical tests carried out on the six-state LR Potts chain with $\sigma=0.7$ show that $\beta(E)$ varies smoothly between the energy peaks of the ordered and disordered phases, which ensures that $\meanval{W_{flip}}$ remains close to 1. 
The variance of $\Delta(E)$ increases whenever $E$ lies outside the range of phase coexistence, and, as is clearly visible, leads to a reduction in the acceptance rate. 
Still and all, it is worth underlining that the energy range of interest in the analysis of first-order phase transitions spans an interval which is only moderately larger than the one corresponding to phase coexistence. In this respect, a mean acceptance rate remaining well above $90\%$ inside this range of energy represents already an improvement of a factor $3$ with respect to the standard multicanonical approach where usual acceptance rates hardly exceed $30\%$ \cite{ReynalDiep2004a}, let alone the crucial fact that a whole lattice sweep is now carried out in a single step.

With regards to performance measurements at first-order transitions, tunneling times have so far been considered one of the most meaningful measurement parameters \cite{JankeSauer1994,Janke2002proc,JankeKappler1995b}. They are defined as one half of the average number of Monte Carlo update sweeps needed for the walk to travel from one peak of the energy histogram to the other -- where peaks are defined with respect to the finite-size transition temperature -- and turn out to represent a good indicator of the interval between roughly independent samples.
\begin{figure}
	\centering
	\InsertFig{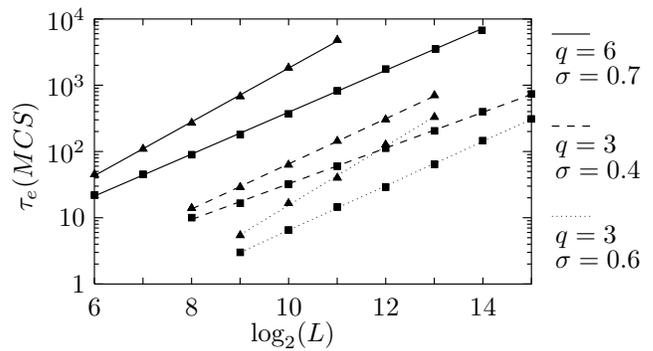}
	\caption{Tunneling times for the long-ranged Potts chain. Triangles and squares refer to the local- and collective-update algorithm, respectively.
	}				
	\label{fig:dyn_exp_potts1D}
\end{figure}
Results for the LR model with $q=3$ and $6$ are shown in Fig.~\ref{fig:dyn_exp_potts1D}.
Dynamic exponents $z$ were determined from a fit to the power law $\tau_e \sim L^{z}$. The collective-update algorithm yields $z=0.89(1)$ and $z=1.11(1)$  for $q=3$, $\sigma=0.4$ and $\sigma=0.6$, and $z=1.05(1)$ for $q=6$, $\sigma=0.7$. This represents a substantial reduction with respect to the local-update implementation where we obtained, respectively, $z=1.13(2)$, $z=1.48(2)$ and $z=1.35(3)$; the improvement is even higher when one consider prefactors.
For the NN model, simulations were performed over lattice sizes ranging from $16 \times 16$ to $256 \times 256$, giving $z=1.82(2)$ for $q=7$ and $z=2.23(1)$ for $q=10$. These estimates are much closer to the ideal value $z \sim 2$ expected from a random walk argument than those obtained with a local-update algorithm, namely $z=2.60(4)$ and $z=2.87(4)$, respectively. Additionally, they compare extremely well with those obtained with the multibond method \cite{JankeKappler1995b} and with Rummukainen's hybrid-like two-step algorithm \cite{Rummukainen1993}, although these approaches and ours differ markedly in the way clusters are constructed. 

In order to check that our algorithm did not produce systematic errors, we computed transition temperatures and interface tensions between coexisting phases for the NN model, for which exact results exist (\cite{BorgsJanke1992,Buddenoir1993} and references in \cite{Janke2002proc}).
For $q=10$, we obtained $T_c(L)=0.70699(5)$, $0.70300(2)$, $0.70278(1)$, $0.70164(1)$, and $0.701328(4)$ for $L=16$, $30$, $32$, $64$ and $128$, where $T_c$ was determined from the location of peaks of the specific heat. Following standard FSS theory at first-order transitions, we collapsed $C_v(T)/L^2$ vs. $(T-T_c)L^2$ over the four highest lattice sizes and found an infinite size temperature $T_c(\infty)=0.70123(5)$ in perfect agreement with the exact value $0.7012315\ldots$ The same procedure applied to $q=7$ and $L=32$, $64$, $128$ and $256$ yielded $T_c(\infty)=0.77306(1)$ which again matches perfectly the exact value $0.7730589\ldots$
We estimated the interface tension $\Sigma$ from the histogram of the energy reweighted at a temperature where energy peaks have the same height, namely, $2\Sigma = -L^{-1}\ln(P_{min})$, where $P_{min}$ denotes the minimum of the histogram between the two energy peaks, and the peak heights are normalized to unity.
Our algorithm allowed us to determine $\Sigma$ with a four-digit precision for sizes up to $L=256$ and nonetheless rather modest statistics (of order $10^7$ sweeps). For the seven-state NN model, we obtained $2\Sigma=0.0336(6)$, $0.0294(1)$, $0.02631(8)$ and $ 0.02384(9)$ for $L=32$, $64$, $128$ and $256$; a linear fit of the form $\Sigma \sim \Sigma(\infty)+c/L$ \cite{LeeKosterlitz1991} performed over the three largest sizes (i.e., for $L$ above the disordered phase correlation length $\xi\sim 48$ \cite{Buddenoir1993}) yielded the infinite size value $0.02230(11)$, still above the exact value $0.020792$, yet closer to it than estimates reported in several previous studies \cite{Rummukainen1993,JankeKappler1995b,Janke2002proc}. We note in passing that this discrepancy may be very well attributed to the influence of higher order terms in the vicinity of $L\sim\xi$, since retaining the two largest sizes only would yield a closer value of $0.02137(20)$.

\begin{figure}
	\centering
	\InsertFig{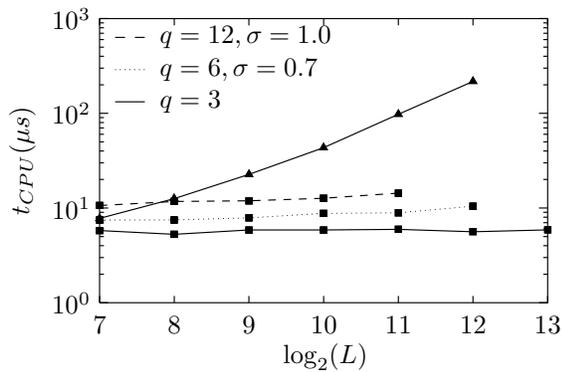}
	\caption{CPU time per MC step and per spin for the long-ranged Potts chain.
	Triangles indicate typical CPU times for the local-update algorithm, irrespective of $q$ and $\sigma$.
	Filled squares refer to the collective-update algorithm, where for $q=3$ estimates were determined 
	by averaging over $\sigma=0.4$, $0.5$ and $0.6$.}
	\label{fig:cpu_time_1D}
\end{figure}
Finally, we discuss CPU demand performances in the case of LR models. Assuming a decently efficient algorithm implementation, this indicator gives a rough account of the algorithm complexity.
Figure~\ref{fig:cpu_time_1D} sketches averages of the CPU (user) time per MC step and per spin. Small fluctuations might be attributed to the effect of CPU caches differing in size. While for the local-update implementation the demand in CPU per spin grows linearly with the number of spins, it is roughly constant over a fairly large range of lattice sizes with the collective-update algorithm. Moreover, the local-update implementation is outperformed already at sizes of several hundreds spins, with nonetheless an increased footprint for higher $q$ due to the correspondingly higher number of FFT's to be computed. This clearly demonstrates the breakthrough that this new method brings about for the numerical study of LR models, drawing in particular highly precise tests of finite-size scaling within computation range.

Our method is easily generalized to other spin models for which a cluster representation is available. Amongst promising candidates are disordered models, which are known to exhibit rugged energy landscapes and high dynamic exponents; related implementation details and results are reported in a distinct work \cite{ReynalDiep2004d}.

%\bibliography{these}

\end{document}